\newcommand{\CLASSINPUTtoptextmargin}{0.75in}%
\newcommand{\CLASSINPUTbottomtextmargin}{1in}%
\newcommand{\CLASSINPUTinnersidemargin}{0.63in}%
\newcommand{\CLASSINPUToutersidemargin}{0.63in}%
\documentclass[conference,10pt,letterpaper]{IEEEtran}% 
\usepackage{amsmath}% for double integral symbol in this template
\usepackage{graphicx}% for figures
\usepackage{multirow}% to allow multiple-row elements in tabular environment
\usepackage[none]{hyphenat}% turn off hyphenation to make text extraction and indexing easier
\usepackage{float}% better control of floating figures and tables
\usepackage{subfig}% for subfigures within figures
\usepackage{dblfloatfix}% fix to allow page-wide floats at bottom of page
\usepackage{t1enc}% allows access to various special characters
\usepackage{times}% change font to Nimbus Roman (based on Times Roman)
\usepackage{threeparttable}
\makeatletter

\def\@IMSauthorblockNAMEstyle{\normalfont\IMSauthorsize}
\def\@IMSauthorblockAFFILstyle{\normalfont\IMSaffilsize}
\def\@IMSauthorblockEMAILstyle{\normalfont\IMSaffilsize}
\def\IMSauthorblockNAME#1{%
\relax\@IMSauthorblockNAMEstyle%
#1%
}%
\def\IMSauthorblockAFFIL#1{%
\relax\@IMSauthorblockAFFILstyle%
\vskip\@IEEEauthorblockAtopspace
#1%
}%
\def\IMSauthorblockEMAIL#1{%
\relax\@IMSauthorblockEMAILstyle%
\vskip\@IEEEauthorblockAtopspace
#1%
}%
\newcommand{\IMSauthor}[1]{%
\ifIsBlindReviewVersion%
\author{\phantom{\parbox{\textwidth}{\center\relax#1}}}%
\else%
\author{\parbox{\textwidth}{\center\relax#1}}%
\fi%
}%
\newif\ifIsBlindReviewVersion
\def\IMSthispaperforblindreview{\IsBlindReviewVersiontrue}
\def\IMSthispaperforfinalpublication{\IsBlindReviewVersionfalse}
\def\@maketitle{\newpage
\bgroup\par\addvspace{0.5\baselineskip}\centering%
\ifCLASSOPTIONtechnote% technotes
   {\bfseries\large\@IEEEcompsoconly{\sffamily}\@title\par}\vskip 1.3em{\lineskip .5em\@IEEEcompsoconly{\sffamily}\@author
   \@IEEEspecialpapernotice\par{\@IEEEcompsoconly{\vskip 1.5em\relax
   \@IEEEtitleabstractindextextbox{\@IEEEtitleabstractindextext}\par
   \hfill\@IEEEcompsocdiamondline\hfill\hbox{}\par}}}\relax
\else% not a technote
   \vskip0.2em{\IMStitlesize\ifCLASSOPTIONtransmag\bfseries\LARGE\fi\@IEEEcompsoconly{\sffamily}\@IEEEcompsocconfonly{\normalfont\normalsize\vskip 2\@IEEEnormalsizeunitybaselineskip
   \bfseries\Large}\@title\par}\vskip1.0em\par% CAUSAL PRODUCTIONS change on this line
   \ifCLASSOPTIONconference%
      {\@IEEEspecialpapernotice\mbox{}\vskip\@IEEEauthorblockconfadjspace%
       \mbox{}\hfill\begin{@IEEEauthorhalign}\@author\end{@IEEEauthorhalign}\hfill\mbox{}\par}\relax
   \else% peerreviewca, peerreview or journal
      \ifCLASSOPTIONpeerreviewca
         {\@IEEEcompsoconly{\sffamily}\@IEEEspecialpapernotice\mbox{}\vskip\@IEEEauthorblockconfadjspace%
          \mbox{}\hfill\begin{@IEEEauthorhalign}\@author\end{@IEEEauthorhalign}\hfill\mbox{}\par
          {\@IEEEcompsoconly{\vskip 1.5em\relax
           \@IEEEtitleabstractindextextbox{\@IEEEtitleabstractindextext}\par\hfill
           \@IEEEcompsocdiamondline\hfill\hbox{}\par}}}\relax
      \else% journal, peerreview or transmag
         \ifCLASSOPTIONtransmag
           {\@IEEEspecialpapernotice\mbox{}\vskip\@IEEEauthorblockconfadjspace%
            \mbox{}\hfill\begin{@IEEEauthorhalign}\@author\end{@IEEEauthorhalign}\hfill\mbox{}\par
           {\vspace{0.5\baselineskip}\relax\@IEEEtitleabstractindextextbox{\@IEEEtitleabstractindextext}\vspace{-1\baselineskip}\par}}\relax
         \else% journal or peerreview
           {\lineskip.5em\@IEEEcompsoconly{\sffamily}\sublargesize\@author\@IEEEspecialpapernotice\par
           {\@IEEEcompsoconly{\vskip 1.5em\relax
            \@IEEEtitleabstractindextextbox{\@IEEEtitleabstractindextext}\par\hfill
            \@IEEEcompsocdiamondline\hfill\hbox{}\par}}}\relax
         \fi
      \fi
   \fi
\fi\par\addvspace{0.0\baselineskip}\egroup}% CAUSAL PRODUCTIONS change on this line, reduce the vspace from 0.5\baselineskip to 0.0

\def\IMStitlesize{\@setfontsize{\IMStitlesize}{18}{21pt}}% CAUSAL PRODUCTIONS change on this line
\def\IMSauthorsize{\@setfontsize{\IMSauthorsize}{12}{13pt}}% CAUSAL PRODUCTIONS change on this line
\def\IMSaffilsize{\@setfontsize{\IMSaffilsize}{12}{13pt}}% CAUSAL PRODUCTIONS change on this line
\def\IMScaptionsize{\@setfontsize{\IMScaptionsize}{8}{9pt}}% CAUSAL PRODUCTIONS change on this line
\def\IMSbibsize{\@setfontsize{\IMSbibsize}{8}{9pt}}% CAUSAL PRODUCTIONS change on this line

\def\@IEEEauthorblockNstyle{\IMSauthorsize\@IEEEcompsocnotconfonly{\sffamily}\@IEEEcompsocconfonly{\large}}%CAUSAL PRODUCTIONS removed sublargesize to get correct IMSauthorsize
\def\@IEEEauthorblockAstyle{\IMSaffilsize\@IEEEcompsocnotconfonly{\sffamily}\@IEEEcompsocconfonly{\itshape}\@IEEEcompsocconfonly{\large}}%CAUSAL PRODUCTIONS removed normalsize to get correct IMSaffilsize
\def\@IEEEauthordefaulttextstyle{\IMSauthorsize\@IEEEcompsocnotconfonly{\sffamily}\sublargesize}%CAUSAL PRODUCTIONS

\def\thebibliography#1{\section*{\refname}%
    \addcontentsline{toc}{section}{\refname}%
    \IMSbibsize\@IEEEcompsocconfonly{\small}\vskip 0.3\baselineskip plus 0.1\baselineskip minus 0.1\baselineskip% CAUSAL PRODUCTIONS change on this line
    \list{\@biblabel{\@arabic\c@enumiv}}%
    {\settowidth\labelwidth{\@biblabel{#1}}%
    \leftmargin\labelwidth
    \advance\leftmargin\labelsep\relax
    \itemsep \IEEEbibitemsep\relax
    \usecounter{enumiv}%
    \let\p@enumiv\@empty
    \renewcommand\theenumiv{\@arabic\c@enumiv}}%
    \let\@IEEElatexbibitem\bibitem%
    \def\bibitem{\@IEEEbibitemprefix\@IEEElatexbibitem}%
\def\newblock{\hskip .11em plus .33em minus .07em}%
\ifCLASSOPTIONtechnote\sloppy\clubpenalty4000\widowpenalty4000\interlinepenalty100%
\else\sloppy\clubpenalty4000\widowpenalty4000\interlinepenalty500\fi%
    \sfcode`\.=1000\relax}

\long\def\@makecaption#1#2{%
\ifx\@captype\@IEEEtablestring%
\par\@IEEEtabletopskipstrut% strut used to align table caption with facing column
\else
\@IEEEfigurecaptionsepspace
\fi
\setbox\@tempboxa\hbox{\normalfont\IMScaptionsize {#1.}\nobreakspace\nobreakspace #2}%
\ifdim \wd\@tempboxa >\hsize%
\setbox\@tempboxa\hbox{\normalfont\IMScaptionsize {#1.}\nobreakspace\nobreakspace}%
\parbox[t]{\hsize}{\normalfont\IMScaptionsize\noindent\unhbox\@tempboxa#2}%
\else
\ifCLASSOPTIONconference \hbox to\hsize{\normalfont\IMScaptionsize\hfil\box\@tempboxa\hfil}%
\else \hbox to\hsize{\normalfont\IMScaptionsize\box\@tempboxa\hfil}%
\fi\fi
\ifx\@captype\@IEEEtablestring%
\@IEEEtablecaptionsepspace
\else
\fi}

\newlength\tablecaptiontotableskip
\newlength\figuretocaptionskip
\def\@IEEEfigurecaptionsepspace{\vskip\figuretocaptionskip\relax}%
\def\@IEEEtablecaptionsepspace{\vskip\tablecaptiontotableskip\relax}%

\def\abstract{\normalfont%
\@IEEEabskeysecsize\bfseries\textit{\abstractname}\,\bfseries\textit{---}\,%
\@IEEEgobbleleadPARNLSP}%

\def\IEEEkeywords{\normalfont%
\@IEEEabskeysecsize\bfseries\textit{\IEEEkeywordsname}\,\bfseries\textit{---}\,%
\@IEEEgobbleleadPARNLSP}%
\def\endIEEEkeywords{\relax\vspace{0.67ex}%
\par\if@twocolumn\else\endquotation\fi%
\normalsize\normalfont}%

\DeclareRobustCommand*{\IMSauthorrefmark}[1]{\raisebox{0pt}[0pt][0pt]{\textsuperscript{\footnotesize{#1}}}}%
\def\@IEEEauthorblockNtopspace{0ex}
\def\@IEEEauthorblockAtopspace{1mm}
\def\tablename{Table}% IMS: was TABLE in IEEETran, but we are now using capitalized caption text not smallcaps
\def\thetable{\arabic{table}}% IMS: Tables are numbered using arabic not roman
\def\IEEEkeywordsname{Keywords}% use Keywords instead of Index Terms
\def\subsubsection{\@startsection{subsubsection}{3}{\z@}{1.5ex plus 1.5ex minus 0.5ex}%
{0.7ex plus .5ex minus 0ex}{\normalfont\normalsize\itshape}}%
\def\@seccntformat#1{\csname the#1dis\endcsname\relax}% moved the spacer \hskip 0.5em to individual handlers below
\def\thesectiondis{\thesection.\hskip 0.5em}%					I.	% CAUSAL PRODUCTIONS: moved the hskip spacer from \@seccntformat to here
\def\thesubsectiondis{{\hbox to\parindent{\Alph{subsection}.}}}%		B.	% CAUSAL PRODUCTIONS: indent the subsection name to match paragraph indent
\def\thesubsubsectiondis{{\hbox to \parindent{\arabic{subsubsection})}}}%	3)	% CAUSAL PRODUCTIONS: indent the subsubsection name to match paragraph indent
\def\theparagraphdis{{\hbox to \parindent{\alph{paragraph})}}}%			d)	% CAUSAL PRODUCTIONS: indent the subsubsubsection name to match paragraph indent
\IEEEilabelindentA \parindent
\IEEEilabelindent \IEEEilabelindentA
\IEEEelabelindent \parindent
\IEEEdlabelindent \parindent
\IEEElabelindent \parindent
\newlength\@IMSparindent
\newcommand\IMSdisplayacksection[1]{%
\ifIsBlindReviewVersion%
\noindent\phantom{\parbox[t]{\columnwidth}{\normalbaselines\setlength{\parindent}{\@IMSparindent}{#1}\strut}}%\IMSacktext
\else%
\noindent\parbox[t]{\columnwidth}{\normalbaselines\setlength{\parindent}{\@IMSparindent}{#1}\strut}%
\fi%
}%

\makeatother

\usepackage{lipsum}
\usepackage{fancyhdr}
\fancypagestyle{sec}{\lhead{\tmpx}}

\fancypagestyle{arXiv}
{
   \fancyhf{}
   \chead{This paper has been accepted to be presented at the 2025 IEEE MTT-S International Microwave Symposium (IMS), \\San Francisco, CA, USA, June 15–20, 2025}
   \cfoot{ \fontsize{=9pt}{10pt}\selectfont  © 2025 IEEE. Personal use of this material is permitted. Permission from IEEE must be obtained for all other uses, in any current or future media, including reprinting/republishing this material for advertising or promotional purposes, creating new collective works, for resale or redistribution to servers or lists, or reuse of any copyrighted component of this work in other works\\\fontsize{=10pt}{12pt}\selectfont\thepage}
   
}

\begin{document}
%%%%%%%%%%%%%%%%%%%%%%%%%%%%%%%%%%%%%%%%%%%%%%%%%%%%%%%%%%%%%%%%%%%%%%%%%%%%%
% We use \raggedbottom to avoid latex adding vertical space around headings.
% This gives a better idea to the author about how much white space remains
% as the page limit is approached.
\raggedbottom
%
%%%%%%%%%%%%%%%%%%%%%%%%%%%%%%%%%%%%%%%%%%%%%%%%%%%%%%%%%%%%%%%%%%%%%%%%%%%%%
% PAPER TITLE AND AUTHOR BLOCK
%
% The paper title can use linebreaks \\ within to get better formatting if desired.
%
\title{TCN-DPD: Parameter-Efficient Temporal Convolutional Networks for Wideband Digital Predistortion}
%
% Next we define the author names and affiliations.
% Author names are listed using \IMSauthorblockNAME{} with comma separators between names.
% Affiliations are listed using \IMSauthorblockAFFIL{} with \\ separators between affiliations.
% Email addresses are listed using \IMSauthorblockEMAIL{} with comma separators between emails.
% See below for examples of each of these.
%
% Symbols marking author-affiliation relations are output using \IMSauthorrefmark{}.
%
% Next we typeset the authorblock either as visible text, or as an empty
% box of the same size, based on the value of the Blind Review Flag.
% Note that the Blind Review Flag also determines whether the Acknowledgments
% section is visible or invisible.
% To set the flag to Blind Review mode, simply uncomment the next line
\IMSthispaperforblindreview
% or to set the flag to Final Paper mode (with author block visible) then
% simply uncomment the next line:
\IMSthispaperforfinalpublication
\IMSauthor{%
\IMSauthorblockNAME{% Author Names
Huanqiang Duan\IMSauthorrefmark{\#},
Manno Versluis\IMSauthorrefmark{\#},
Qinyu Chen\IMSauthorrefmark{\$},
Leo~C.~N.~de~Vreede\IMSauthorrefmark{\#}, 
Chang~Gao\IMSauthorrefmark{\#1}
}% end of \IMSauthorblockNAME
\\%
\IMSauthorblockAFFIL{% Author Affiliations
\IMSauthorrefmark{\#}Department of Microelectronics, Delft University of Technology, The Netherlands\\
\IMSauthorrefmark{\$}Leiden Institute of Advanced Computer Science (LIACS), Leiden University, The Netherlands
}% end of \IMSauthorblockAFFIL
\\%
\IMSauthorblockEMAIL{% Author Emails
\IMSauthorrefmark{1}chang.gao@tudelft.nl
}% end of \IMSauthorblockEMAIL
}% end of \IMSauthor
%
% Next we make the title/author block using the information defined above.
\maketitle
%
%%%%%%%%%%%%%%%%%%%%%%%%%%%%%%%%%%%%%%%%%%%%%%%%%%%%%%%%%%%%%%%%%%%%%%%%%%%%%
% ABSTRACT paragraph.
% in the abstract paragraph.
%
\begin{abstract}
% Digital predistortion (DPD) is essential for mitigating nonlinearity in radio frequency (RF) power amplifiers, particularly for wideband applications. This paper presents TCN-DPD, a novel parameter-efficient architecture based on temporal convolutional networks. By integrating noncausal dilated convolutions with optimized activation functions, our approach achieves superior linearization performance while using significantly fewer parameters than existing deep neural network solutions. Evaluated on the \texttt{OpenDPD} framework with the \texttt{DPA\_200MHz} dataset, TCN-DPD demonstrates superior linearization performance with only 500 real-valued parameters, achieving simulated ACPRs of -51.58/-49.26 dBc (L/R), EVM of -47.52\,dB, and NMSE of -44.61\,dB. These simulation results establish TCN-DPD as a promising solution for efficient wideband PA linearization, with ongoing work focused on hardware validation to confirm real-world performance.
Digital predistortion (DPD) is essential for mitigating nonlinearity in RF power amplifiers, particularly for wideband applications. This paper presents TCN-DPD, a parameter-efficient architecture based on temporal convolutional networks, integrating noncausal dilated convolutions with optimized activation functions. Evaluated on the \texttt{OpenDPD} framework with the \texttt{DPA\_200MHz} dataset, TCN-DPD achieves simulated ACPRs of -51.58/-49.26 dBc (L/R), EVM of -47.52\,dB, and NMSE of -44.61\,dB with 500 parameters and maintain superior linearization than prior models down to 200 parameters, making it promising for efficient wideband PA linearization.
\end{abstract}
%
%%%%%%%%%%%%%%%%%%%%%%%%%%%%%%%%%%%%%%%%%%%%%%%%%%%%%%%%%%%%%%%%%%%%%%%%%%%%%
% THE REST OF THE PAPER follows.
%

\section{Introduction}
\thispagestyle{arXiv}

The increasing demand for high-efficiency, wideband communication systems has intensified the need for effective power amplifier (PA) linearization techniques. As modern wireless systems push toward broader bandwidths and higher efficiency requirements, digital predistortion (DPD) has become indispensable for maintaining signal quality while allowing PAs to operate in their efficient nonlinear regions~\cite{guan2014green}.

The Generalized Memory Polynomial (GMP) model has long been the industry standard for PA linearization~\cite{1703853}. However, as communication bandwidths expand and modulation schemes grow more complex, GMP's limitations in modeling sophisticated memory effects have become increasingly apparent~\cite{hongyo2019deep,lu2024low}. This challenge has driven the exploration of more advanced modeling approaches capable of capturing intricate PA behavior across wider bandwidths.

% Deep learning has emerged as a promising solution for next-generation DPD systems. The evolution began with Time Delay Neural Networks (TDNNs), which demonstrated remarkable effectiveness in modeling the sequential memory characteristics inherent in PAs~\cite{5340581}. This success led to the development of enhanced variants such as the Phase-Normalized Real-valued Time Delay (PN-TDNN) architecture~\cite{fischer2023phase}. The field subsequently witnessed the implementation of various recurrent neural architectures, including Long Short-Term Memory (LSTM)~\cite{hochreiter1997long}, Vector Decomposed Long Short-Term Memory (VDLSTM)~\cite{li2020vector}, and Gated Recurrent Unit (GRU)~\cite{cho2014learning} networks. These architectures have shown exceptional capabilities in capturing temporal dependencies within RF signals. Recent innovations have also incorporated convolutional neural networks (CNNs), exemplified by the RVTDCNN model, which leverages convolutional layers to simultaneously model both spatial and temporal signal characteristics~\cite{hu2021convolutional}.

Deep learning has emerged as a promising solution for next-generation DPD systems. The field evolved from Time Delay Neural Networks (TDNNs)~\cite{5340581} to enhanced variants like Phase-Normalized Real-valued Time Delay (PN-TDNN)~\cite{fischer2023phase}, followed by various recurrent architectures including LSTM~\cite{hochreiter1997long}, VDLSTM~\cite{li2020vector}, and GRU~\cite{cho2014learning}. Recent innovations incorporate convolutional neural networks (CNNs), such as the RVTDCNN model~\cite{hu2021convolutional}, which simultaneously models spatial and temporal signal characteristics.

\begin{figure}[t] %!t
\centering
\includegraphics[width=\linewidth]{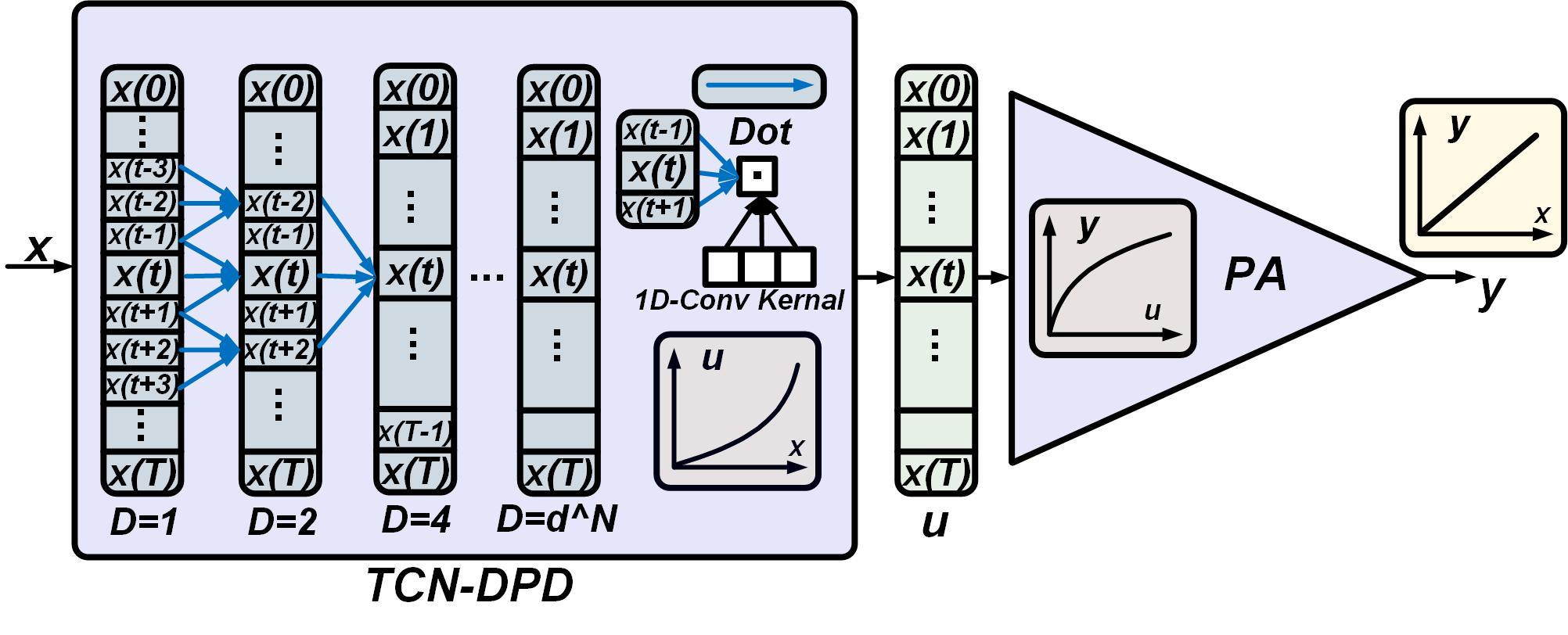}
\caption{Basic concept of TCN-DPD with pre-trained PA model for linearization; example with noncausal 1D convolution layers with dilation sizes $D = 1, 2, 4$ up to $d^N$ in the last layer, and kernel size of 3}
\label{tcn concept}
\end{figure}

Despite these advances, Temporal Convolutional Networks (TCNs) remain largely unexplored in DPD applications, even though they have demonstrated significant advantages in capturing long-range temporal dependencies with reduced computational complexity in time series prediction tasks~\cite{pandey2019tcnn,chen2020probabilistic}. Traditional CNN-based approaches for DPD typically require 2D convolutions operating on framed feature maps extracted from in-phase (I) and quadrature (Q) signals, leading to redundant processing of overlapping time segments. In contrast, TCNs employ 1D convolutions directly on the I/Q time series, combined with dilated convolutional layers that efficiently capture long-term memory effects, as illustrated in Fig.~\ref{tcn concept}. This architectural difference enables TCNs to achieve more parameter-efficient wideband DPD while maintaining or improving modeling accuracy. Importantly, the reduction in model parameters directly translates to lower computational complexity, reduced memory requirements, and ultimately decreased power consumption in practical implementations, a critical consideration for battery-operated devices and energy-efficient base stations.
\setlength{\textfloatsep}{5pt}

This paper proposes TCN-DPD, designed by using noncausal dilated depthwise separable convolutional layers and a residual neural network architecture to achieve superior linearization performance using fewer parameters. As illustrated in Fig.~\ref{tcn concept}, the proposed system pairs a TCN-DPD with a pre-trained PA model.

%-------------------------------------
\section{TCN-DPD Architecture}
%-------------------------------------
\subsection{Noncausal Convolution}
%-------------------------------------
Traditional TCN implementations rely on causal convolution, where network predictions are based exclusively on past data~\cite{pandey2019tcnn,chen2020probabilistic,bai2018empirical}. This design philosophy, inherited from TDNNs, ensures that outputs at time step $t$ depend solely on data from $t$ and earlier timesteps. Such causal systems require leading zero-padding along the time dimension at each layer of TCN to maintain temporal alignment, as defined by Eq.~\ref{equation1}:
\begin{align}
\text{Causal Padding Size} &= K-1
\label{equation1}
\end{align}
where $K$ represents the convolutional kernel size. While one-dimensional fully-convolutional networks (FCN) can partially address this challenge by maintaining consistent input-output sequence lengths~\cite{long2015fully}, they still demand significant computational resources. Our approach implements noncausal convolution, which reduces the required padding size according to Eq.~\ref{equation2}:
\begin{align}
\text{Noncausal Padding Size} &= \frac{K-1}{2} \label{equation2}
\end{align} 
This noncausal design enables the TCN to capture both past and future temporal dependencies within each I/Q data context window of the convolutional kernel size, thereby improving DPD modeling accuracy.
%-------------------------------------
\subsection{Dilated Convolutions}
%-------------------------------------
Sequential tasks requiring extensive historical context often challenge standard convolutional architectures, which struggle to capture long-range temporal dependencies~\cite{bai2018empirical}. The TCN-DPD architecture addresses this limitation by incorporating dilated convolutions, building upon established approaches~\cite{van2016wavenet,yu2015multi}. These dilated convolutions expand the network's receptive field exponentially without requiring additional layers or larger kernels. By progressively increasing dilation sizes across layers, the network efficiently processes extended temporal sequences while maintaining parameter efficiency. Each input within the effective history receives processing from at least one kernel, ensuring comprehensive modeling of long-term dependencies. Fig.~\ref{tcn concept} demonstrates how this dilated structure systematically expands the receptive field across network layers.
%-------------------------------------
\subsection{Proposed DPD Network}
%-------------------------------------
The TCN-DPD architecture, illustrated in Fig.~\ref{tcn}, comprises three main components: input and output 1×1 convolution layers sandwiching a core Depthwise Separable Convolution Block, with activation functions (Act. Funcs.) connecting each layer to enable nonlinear modeling.
\subsubsection{1$\times$1 Convolution}
The network employs 1$\times$1 convolution layers at its input and output boundaries. These layers serve as dimensional adaptation interfaces, efficiently transforming feature representations between the network's internal processing stages.
\subsubsection{Depthwise Separable Convolution Block}
At the architecture's core lies a Depthwise Separable Convolution Block featuring multiple depthwise separable convolution layers~\cite{chollet2017xception}. Each layer implements progressively larger dilation sizes, enabling efficient capture of temporal information through an expanding receptive field. The padding size for these dilated layers follows Eq.~\ref{equation3}:
\begin{align}
\text{Dilated Noncausal Padding Size} &= \frac{K-1}{2} \times D \label{equation3}
\end{align}
where $D$ represents the dilation size and $K$ denotes the kernel size. Each layer incorporates an activation function to model complex nonlinear signal patterns.
%-------------------------------------
\begin{figure}[t] %!t
\centering
\includegraphics[width=\linewidth]{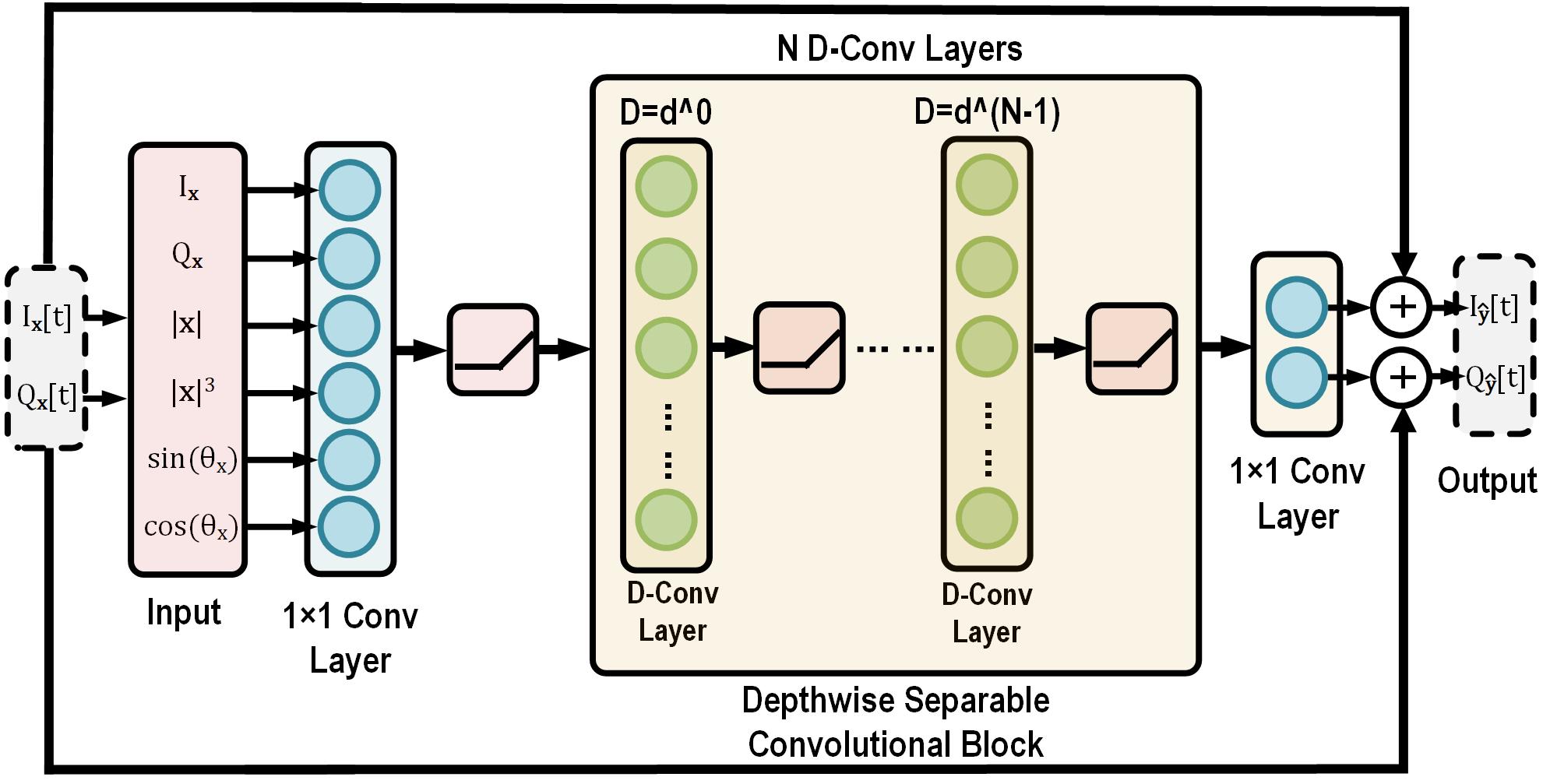}
\caption{The residual TCN architecture with depthwise separable convolutional (D-Conv) layers. $D=d^{N-1}$, where $D$ is the dilation size of each D-Conv layer, $d$ is the dilation base, and $N$ is the number of D-Conv layers.}
\label{tcn}
\end{figure}
%-------------------------------------
\subsubsection{Input Features Extracted from I/Q}
The network processes both I and Q, along with their derived features, as illustrated in the input layer shown in Fig.~\ref{tcn}. This enriched input representation provides comprehensive amplitude and phase information.
\subsubsection{Residual Connections}
The architecture implements residual connections directly linking I and Q inputs to the output, enhancing the model stability in training. These connections preserve critical input information while addressing the vanishing gradient challenges common in deep architectures.

% ==========================
% # III. EXPERIMENTAL RESULTS #
% ==========================

\section{Experimental Results}
\begin{table*}[t]
\centering
\scriptsize % Reduce font size
\begin{threeparttable}
\caption{DPD Performance Comparison of 22 Act. Funcs. Based on a Fixed DGRU PA Pre-trained Model on \texttt{DPA\_200MHz} Validation Set, Averaged Over 5 Random Seeds}

\label{af results}
\begin{tabular}{|c|c|c|c|c|c|c|c|c|c|c|c|}
\hline
ID & 1 & 2 & 3 & 4 & 5 & \textbf{6} & 7 & 8 & 9 & 10 & 11  \\ \hline
Act. Func. & CELU & ELU & GELU & Hardshrink & Hardtanh & \textbf{Hardswish} & LeakyReLU & LogSigmoid & Mish & ReLU & ReLU6  \\ \hline
%SIM-NMSE(dB) & -35.22 & -32.77 & -43.60 & -43.60 & -37.52 & -33.77  & -43.6 & -37.52 & -33.77 & -33.77 & -33.77 \\ \hline  
SIM-NMSE (dB) & -43.62 & -43.62 & -44.79 & -11.10 & -42.00 & \textbf{-44.89}  & -38.20 & -35.62 & -44.33 & -36.74 & -36.77 \\ \hline 
SIM-ACLR (dBc) & -50.21 & -50.20 & -51.52 & -32.92 & -48.00 & \textbf{-51.54}  & -44.16 & -40.88 & -51.17 & -42.53 & -42.53 \\ \hline \hline 
ID & 12 & 13 & 14 & 15 & 16 & 17 & 18 & 19 & 20 & 21 & 22 \\ \hline
Act. Func. & RReLU & SELU & SiLU & Softplus & Softshrink & Softsign & Tanh & Tanhshrink & Hardsigmoid & Sigmoid & PReLU \\ \hline
%SIM-NMSE(dB) & -35.22 & -32.77 & -43.60 & -43.60 & -37.52 & -33.77  & -43.6 & -37.52 & -33.77 & -33.77 & -33.77 \\ \hline  
SIM-NMSE (dB) & -37.25 & -41.08 & -44.70 & -38.59 & -11.10 & -43.65  & -44.58 & -11.10 & -38.05 & -35.62 & -41.70 \\ \hline 
SIM-ACLR (dBc) & -43.06 & -47.53 & -51.54 & -44.02 & -32.92 & -50.14  & -51.37 & -32.92 & -44.88 & -40.88 & -47.98 \\ \hline   
\end{tabular}
\begin{tablenotes} 
\item[a] TCN architecture set up with 4 depthwise convolution layers, kernel size $K=5$, and a predefined dilation base $d=2$.
\end{tablenotes}
\end{threeparttable}
\end{table*}

\begin{figure*}[ht!] %!t
\centering
\includegraphics[width=0.95\linewidth]{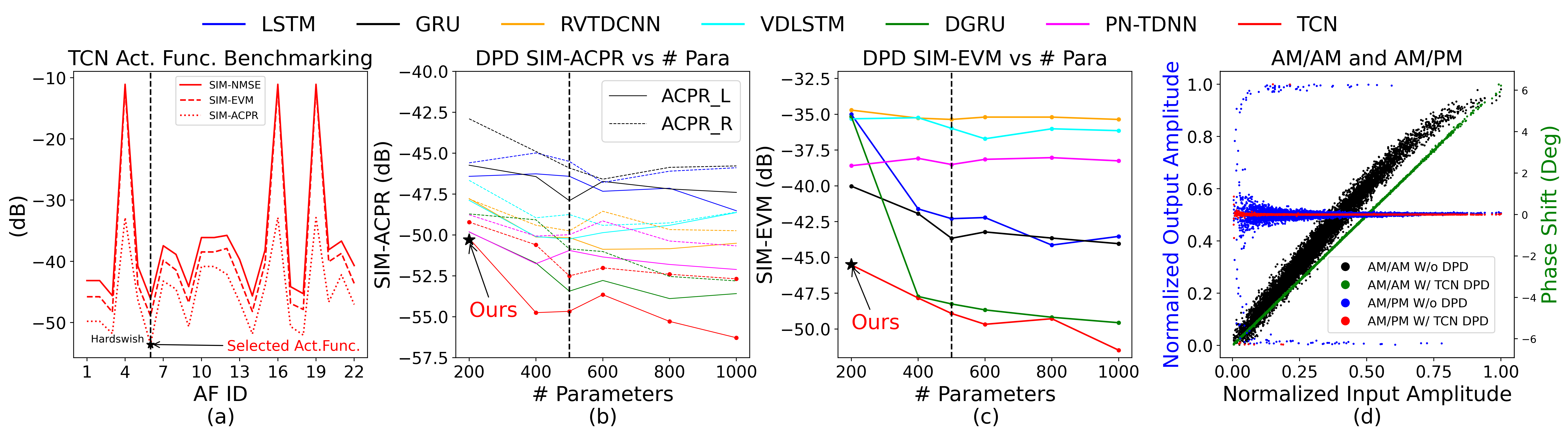}
\caption{Training simulation results on 200 MHz 10-channel $\times$ 20 MHz OFDM signals from the \texttt{DPA\_200MHz} validation set. Each curve represents the best performance of each architecture over 5 random seeds with a fixed DGRU PA model. (a) DPD performance comparison of 22 Act.Funcs. Based on a fixed DGRU PA model. (b) The 500-parameter DPD learning SIM-ACPR\_L/R over training epochs. (c) SIM EVM vs. real-valued model parameters. (d) The 500-parameter DPD modeling AM/AM and AM/PM plot.}
\label{tcn figure}
\end{figure*}

\setlength{\floatsep}{0.5pt} 

\begin{figure}[ht!] %!t
\centering
\includegraphics[width=0.95\linewidth]{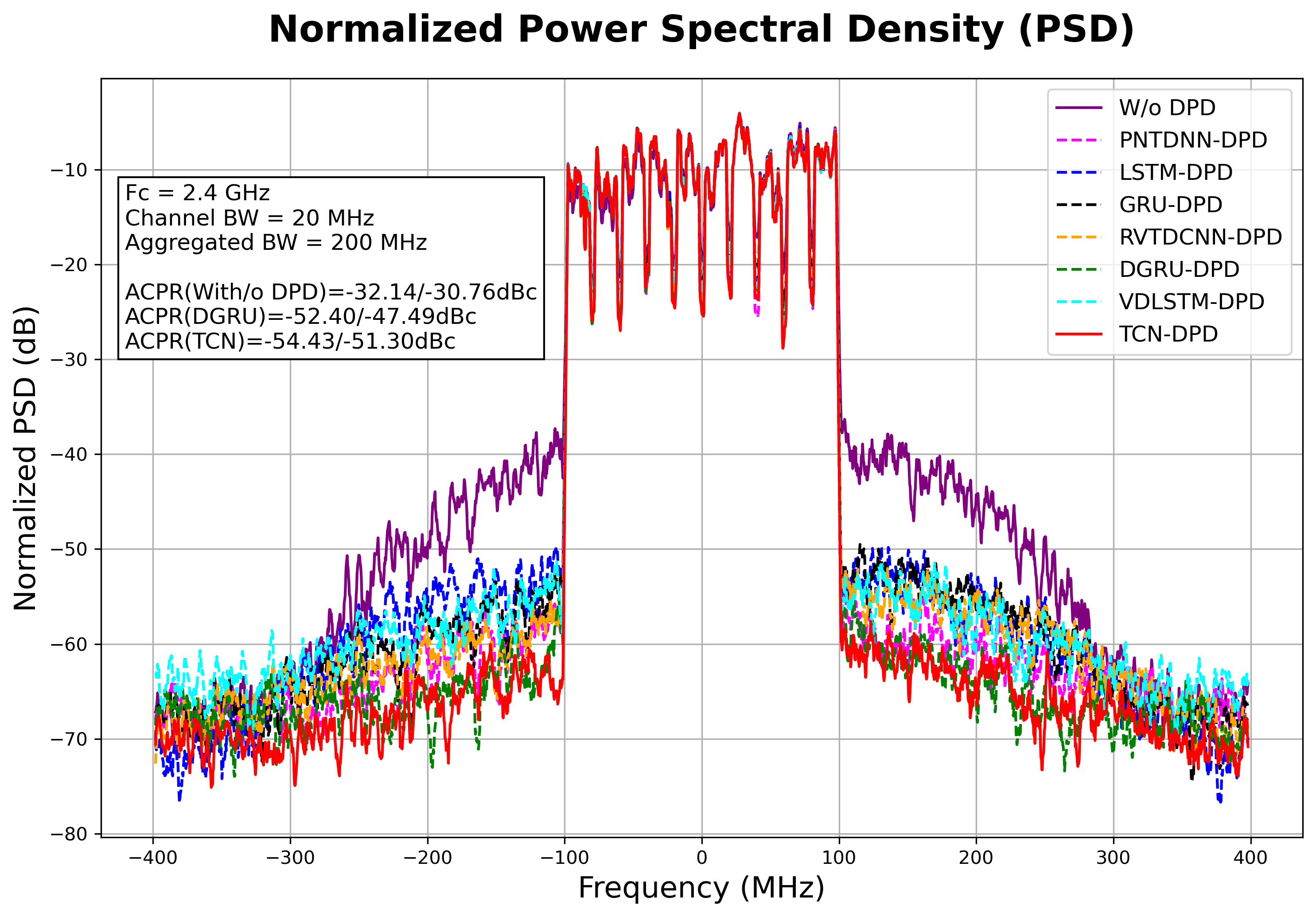}
\caption{PSD comparison of the output signal with different 500 real-valued parameter DNN-DPD models on a 200 MHz signal from the \texttt{DPA\_200MHz} test set. Each curve represents the best DPD performance of each model over 5 random seeds based on a fixed DGRU PA model.}
\label{psd}
\end{figure}

\setlength{\floatsep}{0.5pt} 

\begin{table}[t]
\centering
\scriptsize % Reduce font size

\setlength{\tabcolsep}{2.45pt} % Reduce column spacing
\begin{threeparttable}
\caption{Performance Comparison of DPD Models Based on a Fixed DGRU PA Model with Approximately 500 Real-Valued Parameters on \texttt{DPA\_200MHz} Test Set, Averaged Over 5 Random Seeds $\pm$ Standard Deviations}
\label{tcn results}
\begin{tabular}{|c|c|ccc|}
\hline
\textbf{Classes} & \textbf{DPD Models} 
& \begin{tabular}[c]{@{}c@{}}\textbf{SIM-NMSE} \\ \textbf{(dB)}\end{tabular} 
& \begin{tabular}[c]{@{}c@{}}\textbf{SIM-ACPR} \\ \textbf{(dBc, L/R)}\end{tabular} 
& \begin{tabular}[c]{@{}c@{}}\textbf{SIM-EVM} \\ \textbf{(dB)}\end{tabular}  \\ \hline \hline

W/o DPD\tnote{a} & - & - 
& -31.90$\pm$0.24 / -30.45$\pm$0.31 
& -34.02$\pm$1.42 \\ \hline

\multirow{2}{*}{Base RNN}
& LSTM & -35.22$\pm$3.86 
& -43.60$\pm$1.14 / -42.68$\pm$0.36 
& -37.52$\pm$5.11 \\ 

& GRU & -40.01$\pm$1.68 
& -44.95$\pm$0.65 / -43.76$\pm$1.60 
& -42.70$\pm$2.23  \\ \hline

\multirow{4}{*}{Prior DPD}
& RVTDCNN~\cite{hu2021convolutional} & -32.03$\pm$1.19 
& -48.04$\pm$0.71 / -46.26$\pm$1.27 
& -34.61$\pm$1.81 \\ 

& VDLSTM~\cite{li2020vector} & -32.50$\pm$0.71
& -47.04$\pm$1.44 / -45.85$\pm$1.32
& -34.94$\pm$1.54 \\ 

& PN-TDNN~\cite{fischer2023phase} & -35.49$\pm$0.47
& -49.25$\pm$0.64 / -48.43$\pm$1.11 
& -37.70$\pm$0.89 \\  

& DGRU~\cite{wu2024opendpd} & -41.82$\pm$2.87 
& -50.57$\pm$1.82 / -49.16$\pm$1.67 
& -44.04$\pm$2.38  \\ \hline

\multirow{3}{*}{This Work}
& \textbf{TCN-500} & \textbf{-44.61$\pm$1.37} 
& \textbf{-51.58$\pm$2.84 / -49.26$\pm$2.04} 
& \textbf{-47.52$\pm$1.49}  \\ 

& TCN-200 & -41.27$\pm$1.55
& -45.83$\pm$2.39 / -46.76$\pm$1.23
& -43.81$\pm$1.67 \\ 

& TCN-1000 & -46.37$\pm$1.13
& -52.58$\pm$2.43 / -50.84$\pm$1.44 
& -49.40$\pm$1.90\\ \hline

\end{tabular}
\begin{tablenotes}
\item[a] Based on a fixed DGRU pre-trained PA model over 5 random seeds, the average SIM-NMSE is -31.84 dB.
%\item[b] The input features contain \(I_x\), \(Q_x\), \(|\mathbf{x}|\), \(|\mathbf{x}|^3\), \(\sin\theta_x\), \(\cos\theta_x\).
\end{tablenotes}
\end{threeparttable}
\end{table}
\setlength{\floatsep}{1pt}

\begin{figure}[ht!] %!t
\centering
\includegraphics[width=0.9\linewidth]{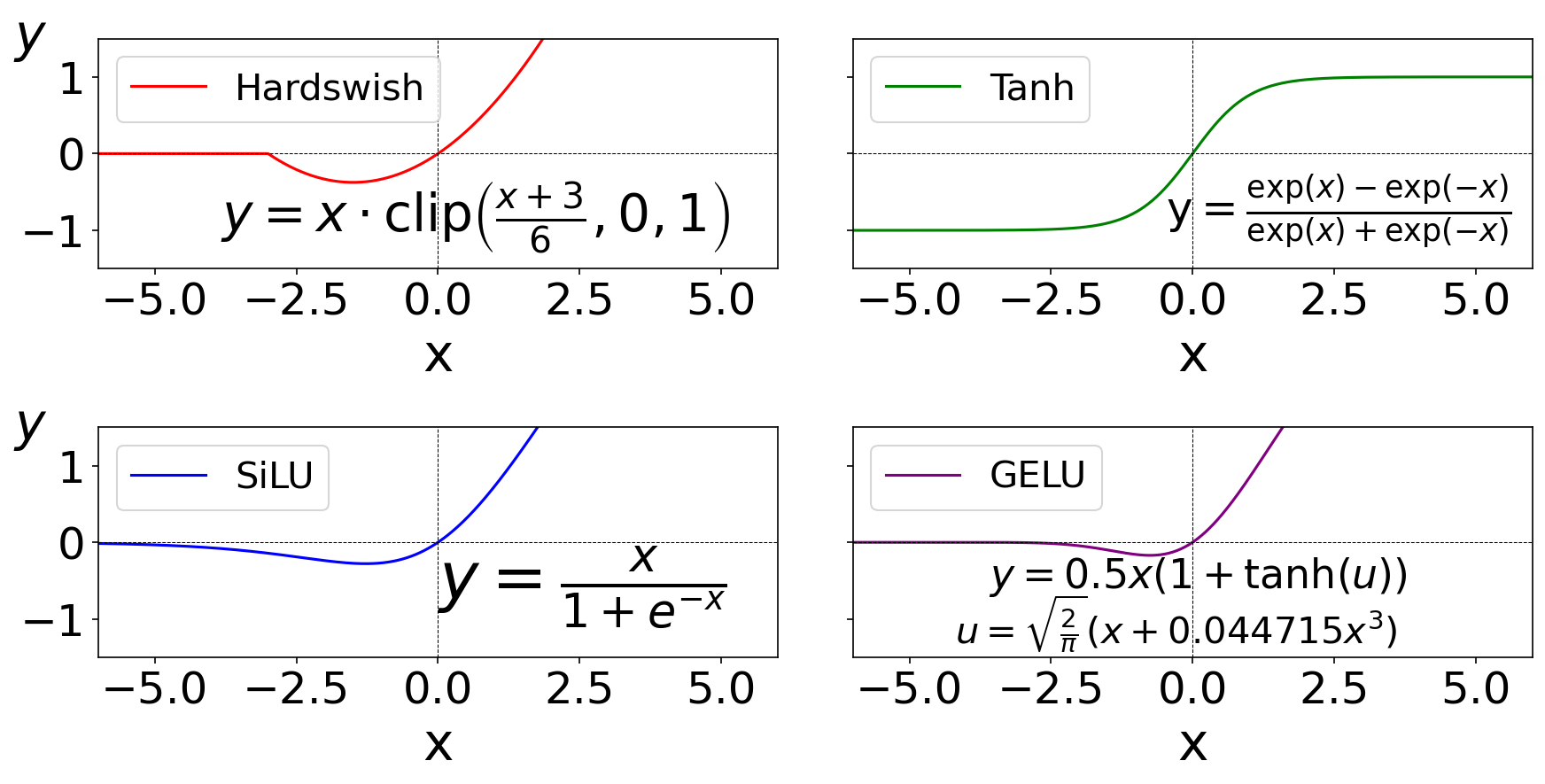}
\caption{The Top-4 performing activation functions in Table~\ref{af results}.}
\label{4 activation}
\end{figure}

\setlength{\textfloatsep}{1pt}

\subsection{Experimental Setup and Benchmarking}
% All experiments were conducted using the \texttt{DPA\_200MHz} dataset~\cite{wu2024opendpd}, which contains 200 MHz bandwidth signals (10 channels × 20 MHz) with 64-QAM OFDM modulation, measured from a 40 nm digital transmitter~\cite{beikmirza2023}. For consistent evaluation, we employed a fixed pre-trained DGRU PA model, achieving an average SIM-NMSE of -31.84 dB, with approximately 500 real-valued parameters tested across 5 random seeds. While simulated results may differ from measured performance, previous research~\cite{wu2024opendpd} has demonstrated that the relative performance rankings among different architectures remain consistent between simulation and measurement, validating our simulation-based evaluation approach. However, we acknowledge that hardware measurements are essential for final validation, which forms a key component of our ongoing work.
All experiments used the \texttt{DPA\_200MHz} dataset in \texttt{OpenDPD}~\cite{wu2024opendpd}, containing 200 MHz bandwidth signals (10 channels × 20 MHz) with 64-QAM OFDM modulation from a 40 nm digital transmitter~\cite{beikmirza2023}. For consistent evaluation, we employed a fixed pre-trained DGRU PA model achieving a SIM-NMSE of -31.84 dB, with approximately 500 real-valued parameters tested across 5 random seeds. While simulation and measurement results may differ, previous research~\cite{wu2024opendpd} showed consistent relative performance rankings between the two approaches. Hardware validation remains essential for final confirmation and is ongoing.

\subsubsection{Benchmark of Activation Functions}
After establishing the core TCN architecture, we systematically investigated activation functions to optimize network performance. Traditional TCN implementations often employ ReLU or its variants~\cite{pandey2019tcnn, bai2018empirical}. However, these functions present two significant limitations for DPD applications: their inability to process negative values effectively and their non-smooth characteristics, which can introduce spectral leakage. These limitations typically require additional network parameters to model the target function accurately.

To identify the most effective activation function, we evaluated 22 different options using a baseline TCN configuration with 4 depthwise convolution layers, kernel size $K=5$, and dilation base $d=2$. Each activation function was tested across 5 random seeds, with results presented in Fig.~\ref{tcn figure}(a). To ensure reliability, we computed averaged performance metrics across all seeds, as detailed in Table~\ref{af results}.

\subsubsection{DPD Benchmarking}
Following activation function optimization, we conducted comprehensive performance benchmarking of the TCN-DPD architecture. We evaluated configurations ranging from 200 to 1000 parameters against other state-of-the-art DNN-DPD models. Each configuration underwent testing across 5 random seeds to ensure statistical significance. The best-performing models for each parameter count are presented in Fig.~\ref{tcn figure}(b) and (c). In contrast, Table~\ref{tcn results} provides detailed performance metrics for the 500-parameter configuration, averaged across all random seeds to demonstrate robustness.

\subsection{Results and Discussion}
% Our analysis reveals significant insights regarding both activation function selection and overall model performance. The comparative evaluation of activation functions, presented in Fig.~\ref{tcn figure}(a), demonstrates that Hardswish consistently achieves superior performance across all metrics. This finding is further validated by the averaged results in Table~\ref{af results}, establishing Hardswish's robust performance advantage. Three other activation functions, GELU, SiLU, and Tanh, also demonstrate strong performance. As shown in Fig.~\ref{4 activation}, these functions share essential characteristics: they are differentiable across their domains, operate effectively over both positive and negative ranges, and maintain smooth gradient transitions. These properties make them particularly well-suited for DPD applications, offering viable alternatives to the commonly used ReLU variants in deep learning architectures.
Our analysis reveals insights regarding activation function selection and model performance. The evaluation in Fig.~\ref{tcn figure}(a) and Table~\ref{af results} demonstrates that Hardswish consistently achieves superior performance across all metrics. Three other functions, GELU, SiLU, and Tanh, also perform strongly. As shown in Fig.~\ref{4 activation}, these functions share key characteristics: differentiability across their domains and smooth gradient transitions near $x=0$, making them well-suited for DPD applications.

The comprehensive performance evaluation of the TCN-DPD architecture, illustrated in Fig.~\ref{tcn figure}(b) and (c), demonstrates exceptional modeling capabilities across different parameter configurations. Using SIM-ACPR and SIM-EVM as primary metrics, our TCN-based model consistently outperforms existing architectures across parameter ranges from 200 to 1000. Most notably, with just 200 parameters, the TCN achieves remarkable performance with average ACPRs (L/R) of -45.83 and -46.76 dBc and an exceptional average EVM of -43.81 dB. This achievement is particularly significant as competing architectures struggle to maintain comparable EVM performance at such low parameter counts, highlighting the TCN's potential as a highly efficient DPD solution.

The performance of TCN-DPD extends across all parameter configurations, especially with advantages at lower parameter counts. This consistent performance underscores the TCN architecture's fundamental efficiency in capturing PA nonlinearities while maintaining parameter efficiency. Table~\ref{tcn results} presents a detailed performance comparison at the 500-parameter configuration point, where our architecture demonstrates clear advantages over existing approaches. The AM/AM and AM/PM characteristics plotted in Fig.~\ref{tcn figure}(d), derived from the best-performing model over 5 random seeds, further validate the TCN's modeling accuracy. The power spectral density (PSD) comparison in Fig.~\ref{psd} provides clear visual evidence of the TCN's superior linearization capabilities.

% ==================
% # Conclusion #
% ==================
\section{Conclusion}
% This paper introduces TCN-DPD, a novel parameter-efficient DPD architecture for wideband power amplifier linearization. By carefully integrating noncausal dilated convolutions and optimized activation functions, our approach achieves superior simulated ACPR, EVM, and NMSE metrics while using significantly fewer parameters than prior solutions. While our simulation results are promising, we recognize the importance of hardware validation. Our ongoing work includes implementing the TCN-DPD architecture in a hardware testbed to verify its real-world performance advantages, power consumption benefits, and practical implementation considerations. These measurement results will be reported in a future publication.
This paper introduces TCN-DPD, a novel parameter-efficient architecture for wideband power amplifier linearization. By integrating noncausal dilated convolutions with optimized activation functions, our approach achieves superior simulated ACPR, EVM, and NMSE metrics while using significantly fewer parameters than prior solutions. Our ongoing work focuses on hardware implementation to verify real-world performance, with results to be reported in a future publication.

\section*{Acknowledgment}
This work was partially supported by the European Research Executive Agency (REA) under the Marie Skłodowska-Curie Actions (MSCA) Postdoctoral Fellowship program, Grant No. 101107534 (AIRHAR).

%%%%%%%%%%%%%%%%%%%%%%%%%%%%%%%%%%%%%%%%%%%%%%%%%%%%%%%%%%%%%%%%%%%%%%%%%%%%%

\bibliographystyle{IEEEtran}

\bibliography{IEEEabrv,IEEEexample}

\end{document}